\documentclass[preprint,showpacs,aps]{revtex4}
\usepackage{graphicx}% Include figure files

\begin{document}

\title{Discrete models of dislocations and their motion in cubic crystals}

\author{A. Carpio\cite{carpio:email}}

\affiliation{Departamento de Matem\'{a}tica Aplicada, Universidad
Complutense de Madrid, 28040 Madrid, Spain}

\author{ L. L. Bonilla\cite{bonilla:email} }

\affiliation{Grupo de Modelizaci\'on y Simulaci\'on Num\'erica,\\
Escuela Polit\'ecnica Superior,  Universidad Carlos III de Madrid, 
Avenida de la Universidad 30, 28911 Legan{\'e}s, Spain}

\date{ \today  }

\begin{abstract}
A discrete model describing defects in crystal lattices and having the
standard linear anisotropic elasticity as its continuum limit is proposed. The
main ingredients entering the model are the elastic stiffness constants of the
material and a dimensionless periodic function that restores the
translation invariance of the crystal and influences the Peierls stress.
Explicit expressions are given for crystals with cubic symmetry: sc, fcc and bcc.
Numerical simulations of this model with conservative or damped dynamics
illustrate static and moving edge and screw dislocations and describe their
cores and profiles. Dislocation loops and dipoles are also
numerically observed. Cracks can be created and propagated by applying a
sufficient load to a dipole formed by two edge dislocations.
\end{abstract}


\maketitle

\section{Introduction}
The advances of electronic microscopy allow imaging of atoms and can therefore
be used to visualize the core of dislocations \cite{nab67,hir82}, cracks
\cite{fre90} and other defects that control crystal growth and the mechanical,
optical and electronic properties of the resulting materials \cite{science}.
Emerging behavior due to motion and interaction of defects might explain common
but poorly understood phenomena such as friction \cite{ger01}. Defects can be
created in a controlled way by ion bombardment on reconstructed surfaces
\cite{RGR01}, which allows the study of effectively two dimensional (2D) single
dislocations and dislocation dipoles. These dislocations are effectively 2D
because the surface `floats' on the 3D crystal \cite{FMS93}. Other defects that
are very important in multilayer growth are misfit dislocations 
\cite{MBl74,hul01,BRB97}. At the nanoscale, many processes (for example,
dislocation emission around nanoindentations \cite{rojo}) involve the
interaction of a few defects so close to each other that their core structure
plays a fundamental role. To understand them, the traditional method of using
information about the far field of the defects (extracted from linear
elasticity) to infer properties of far apart defects reaches its limits. The
alternative method of {\em ab initio} simulations is very costly and not very
practical at the present time. Thus, it would be interesting to have
systematic models of defect motion in crystals that can be solved cheaply, are
compatible with elasticity and yield useful information about the defect cores
and their mobility.

\begin{figure}
\begin{center}
\includegraphics[width=8cm]{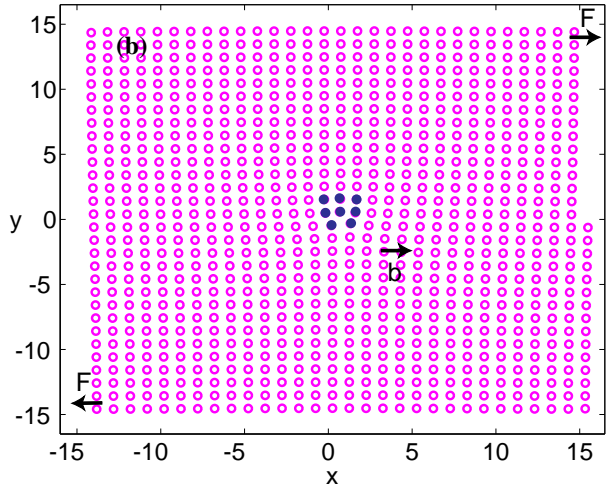}
\caption{(Color online) Deformed cubic lattice in the presence of an edge dislocation 
for the piecewise linear $g(x)$ of Eq.\ (\ref{e9}) with $\alpha=0.24$. }
\label{fig:edge1}
\end{center}
\end{figure}

To see what these models of defects might be like, it is convenient to recall a
few facts about dislocations. Consider for example an edge dislocation in a simple 
cubic (sc) lattice with a Burgers vector equal to one interionic distance in gliding motion, 
as in Fig. \ref{fig:edge1}. The atoms above the $xz$ plane glide over those below. 
Let us label the atoms by their position before the dislocation moves beyond the origin. 
Consider the atoms $(x_{0},-a/2,0)$ and $(x_{0},a/2,0)$  which are nearest 
neighbors before the dislocation passes them. After the passage of the dislocation, 
the nearest neighbor atoms are $(x_{0},-a/2,0)$ and $(x_{0}-a,a/2,0)$. 
This large excursion is incompatible with the main assumption under which
linear elasticity is derived for a crystal structure \cite{ash76}: the deviations
of ions in a crystal lattice from their equilibrium positions are small (compared to the 
interionic distance), and therefore the ionic potentials entering the total potential energy 
of the crystal are approximately harmonic. One obvious way to describe dislocation motion 
is to simulate the atomic motion with the full ionic potentials. This description is possibly 
too costly. In fact, we know that the atomic displacements are small far from the dislocation 
core and that linear elasticity holds there. Is there an intermediate description that allows 
dislocation motion in a crystal structure and is compatible with a far field described by 
the corresponding anisotropic linear elasticity?

If we try to harmonize the continuum description of dislocations according to elasticity
with an discrete description which is simply elasticity with finite differences instead of 
differentials, we face a second difficulty. The displacement vector of a static edge dislocation 
is multivalued. For example, its first component is $\tilde{u}_{1}=a (2\pi)^{-1}[
\tan^{-1}(y/x)+ xy/(2(1-\nu)(x^2+y^2))]$ for the previously described edge dislocation 
($\nu$ is the Poisson ratio) \cite{nab67}. In elasticity, this fact does not cause any trouble 
because the physically relevant strain tensor contains only derivatives of the displacement 
vector. These derivatives are continuous even across the positive $x$ axis, where the 
displacement vector has a jump discontinuity $[u_{1}]=a$. If we consider a discrete model, 
and use differences instead of differentials, the difference of the displacement vector may still 
have a jump discontinuity across the positive $x$ axis. 

The previous difficulties have been solved in a simple discrete model of edge dislocations and
crowdions called the IAC model (interacting atomic chains model) proposed and studied 
by A.I.\ Landau and collaborators \cite{kkl}. A similar model for screw dislocations in
bcc crystals was proposed earlier by H. Suzuki \cite{suzuki}.  In the equations for the IAC 
model, the differences of the displacement vector are replaced by their sines. Unlike the finite
differences, these sine functions are continuous across the positive $x$ axis. Moreover,
the equations remain unchanged if a horizontal chain of atoms slides an integer number of 
lattice periods $a$ over another chain. Taking advantage of its simplicity, we have recently
analyzed pinning and motion of edge dislocations in the IAC model \cite{car03}.

\begin{figure}\begin{center}
\includegraphics[width=8cm]{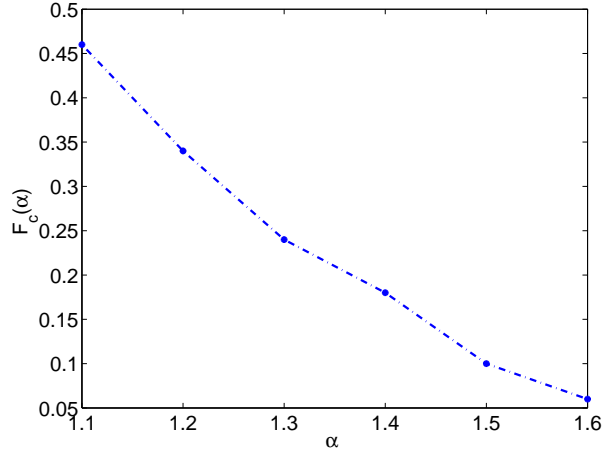}
\end{center}
\caption{(Color online) Peierls stress in dimensionless units for a 2D edge dislocation in 
a sc crystal with the stiffnesses of tungsten as a function of the parameter $\alpha$ in the 
periodic function $g(x)$.}
\label{fcritico}
\end{figure}

In this paper, we propose a top-down approach to discrete models of dislocations in cubic 
crystals. Let us start with a simple cubic lattice having a unit cell of side length $a$. Firstly, 
we discretize space along the primitive vectors defining the unit cell of the crystal: $x=x_{1}
=l a$, $y=x_{2}=ma$, $z=x_{3}= n a$, where $l$, $m$ and $n$ are integer numbers. We 
shall measure the displacement vector in units of $a$, so that $\tilde{u}_i(x,y,z,t)= a\, 
u_i(l,m,n;t)$ and $u_i(l,m,n;t)$ is a nondimensional vector. Let $D^+_j$ and $D^-_j$ 
represent the standard forward and backward difference operators, so that $D^\pm_1 
u_i(l,m,n;t) = \pm\, [u_i(l\pm 1,m,n;t) - u_i(l,m,n;t)]$, and so on. We shall define the {\em 
discrete} distortion 
tensor as
\begin{eqnarray} 
w_{i}^{(j)} = g(D^+_j u_i), \label{e6}
\end{eqnarray} 
where $g(x)$ is a periodic function of period one satisfying $g(x)\sim x$ as $x\to 0$. In this 
paper, we shall use the odd continuous piecewise linear function: 
\begin{equation} 
g(x) = \left\{ \begin{array} {ll}
x , \quad |x|< {1\over 2} - \alpha,\\
{(1 -2\alpha) (1 - 2x)\over 4\alpha}\,,
\quad {1\over 2} - \alpha < x < {1\over 2} , \end{array} \right. 
\label{e9}
\end{equation} 
which is periodically extended outside the interval $(-1/2,1/2)$ for a given $\alpha$, $0<
\alpha<1/2$. Note that $g(x)$ is symmetrical if $\alpha=1/4$ and that the interval of $x$
in which $g'(x)<0$ widens with respect to that in which $g'(x)>0$ as $\alpha$ increases. 
Numerical simulations of the governing equations for a 2D edge dislocation show that the 
Peierls stress decreases as $\alpha$ increases; see Fig.~\ref{fcritico}, which will be further
commented later on. This means that the dislocation is harder to move if $\alpha$ decreases, 
i.e., if the interval of $x$ in which $g'(x)<0$ shrinks with respect to that in which $g'(x)>0$. 
The parameter $\alpha$ can be selected so as to agree with the observed or calculated Peierls 
stress of a given crystal. 

Secondly, we replace the strain tensor in the strain energy by
\begin{eqnarray} 
e_{ij} = {1\over 2}\, (w_{i}^{(j)} + w_{j}^{(i)}) = {g(D^+_j u_i) +
g(D^+_i u_j)\over 2} . \label{e7}
\end{eqnarray} 
Summing the strain energy over all lattice sites, we obtain the potential energy of the crystal:
\begin{eqnarray} 
V(\{u_i\}) &=& a^3\, \sum_{l,m,n} W(l,m,n;t) . \label{e8}\\
W(l,m,n;t) &=& W(\{u_i\})= {1\over 2} c_{ijkl} e_{ij} e_{kl}, \label{e1}\\
c_{ijkl} &=& \lambda\, \delta_{ij} \delta_{kl} + \mu\,
(\delta_{ik} \delta_{jl} + \delta_{il} \delta_{jk})  + 2 (C_{44} - \mu)\nonumber \\ 
&\times& \left( {\delta_{ik} \delta_{jl} + \delta_{il} \delta_{jk}
\over 2} -\delta_{1i} \delta_{1j}\delta_{1k} \delta_{1l}
  - \delta_{2i} \delta_{2j}\delta_{2k} \delta_{2l} -\delta_{3i} \delta_{3j}
\delta_{3k} \delta_{3l} \right) ,
\label{e2}
\end{eqnarray} 
in which summation over repeated indices is understood. Here, $\lambda=C_{12}$, $\mu=
(C_{11}-C_{12})/2$, where $C_{ij}$ are the stiffness constants of a cubic crystal. If 
$C_{44}=\mu$, the strain energy is isotropic and $\lambda$ and $\mu$ are the usual Lam\'e
coefficients. 

Next, we find the equations of motion. In the absence of dissipation and fluctuation effects,
they are 
\begin{eqnarray} 
\rho a^4\, \ddot{u}_{i}(l,m,n;t) = -{1\over a}\, 
{\partial V(\{u_k\})\over \partial u_i(l,m,n;t) } , \label{e10}
\end{eqnarray} 
or, equivalently,
\begin{eqnarray} 
M\, \ddot{u}_{i}(l,m,n;t) &=& -{\partial\over
\partial u_i(l,m,n;t)}\sum_{l',m',n'} W(l',m',n';t) , \label{e11}\\   
M &=& \rho a^2. \label{e12}
\end{eqnarray} 
Here $\ddot{u}_{i} \equiv \partial^2u_{i}/\partial t^2$, $M$ has units of mass per 
unit length (because $\rho$ is the mass density) and the displacement vector is dimensionless, 
so that both sides of Eq.\ (\ref{e11}) have units of force per unit area. We show in 
Appendix \ref{appEqs} that Eq.\ (\ref{e11}) is equivalent to the following spatially
discrete equations
\begin{eqnarray} 
M\, \ddot{u}_{i} &=& \sum_{j,k,l} D^-_j [c_{ijkl}\, g'(D^+_j u_i)\, 
g(D^+_l u_k)].   \label{e13}
\end{eqnarray}
To nondimensionalize these equations, we could adopt $C_{44}$ as the typical scale of stress 
and $t_{0}= \sqrt{M/C_{44}}= a\sqrt{\rho/C_{44}}$ as the time scale. The resulting 
equations are the same ones with $M=1$ and $c_{ijkl}/C_{44}$ instead of $c_{ijkl}$.
Let us now restore dimensional units to this equation, so that $\tilde{u}_i(x,y,z)= a\,
u_i(x/a,y/a,z/a)$, then let $a\to 0$, use Eq.\ (\ref{e12}) and that $g(x)\sim x$ as $x\to 0$.
Then we obtain the equations of linear elasticity \cite{ll7},
\begin{eqnarray} 
\rho\, {\partial^2\tilde{u}_{i}\over\partial t^2} =
 \sum_{j,k,l} {\partial\over\partial x_j}\left(
c_{ijkl}\, {\partial \tilde{u}_k\over\partial x_l}\right). \label{e14}
\end{eqnarray}
Thus the discrete model with conservative dynamics yields the Cauchy equations for
elastic constants with cubic symmetry provided the components of the distortion tensor
are very small (which holds in the dislocation far field). Equations of motion with 
dissipation and fluctuation terms can be obtained by writing a quadratic dissipative 
function which, in the isotropic case, yields the usual fluid viscosity terms and using
the fluctuation-dissipation theorem \cite{unpub}.

In the rest of the paper, we describe dislocation motion in simple cubic crystals and 
extend our discrete elastic equations to the case of fcc and bcc lattices.

\section{Dislocation motion in sc crystals}
\label{sec:dynamics}
In this Section, we shall find numerically pure screw and edge dislocations of our discrete
model for sc symmetry and discuss their motion under appropriate applied stresses. In all 
cases, the procedure to obtain numerically the dislocation from the discrete model equations
is the same. We first solve the stationary equations of continuum elasticity with
appropriate singular source terms to obtain the {\em dimensional} displacement vector 
${\bf \tilde{u}}(x,y,z) = (\tilde{u}_{1}(x,y,z),\tilde{u}_{2}(x,y,z),\tilde{
u}_{3}(x,y,z))$ of the static dislocation {\em under zero applied stress}. This displacement 
vector yields the far field of the corresponding dislocation for the discrete model, which is 
the {\em nondimensional} displacement vector: ${\bf U}(l,m,n)= {\bf\tilde u}(la,ma,
na)/a$. We use the nondimensional static displacement vector ${\bf U}(l,m,n)$ in the 
boundary and initial conditions for the discrete equations of motion of the discrete model. 
Later in the Section, we shall show numerical results corresponding to the interaction of 
edge dislocations and the opening of a crack.

\subsection{Screw dislocations}
\label{sec:screw}
\begin{figure}\begin{center}
\includegraphics[width=8cm]{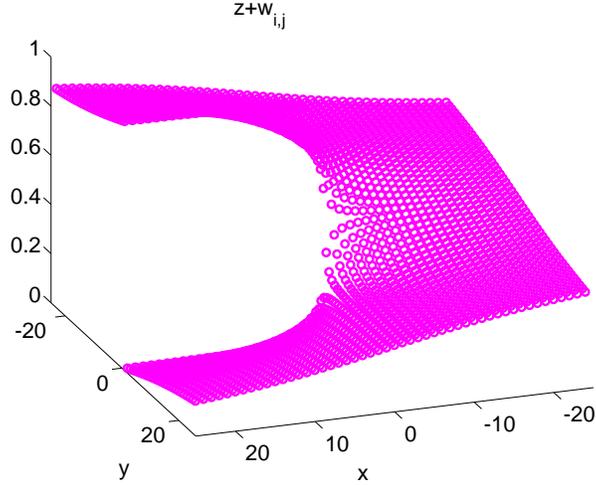} %{helicep1f.eps}
\caption{(Color online) Screw dislocation for the piecewise linear $g(x)$ of Eq.\ 
(\ref{e9}) with $\alpha=0.24$.}
\label{screw1}
\end{center}
\end{figure}

\begin{figure}\begin{center}
\includegraphics[width=7cm]{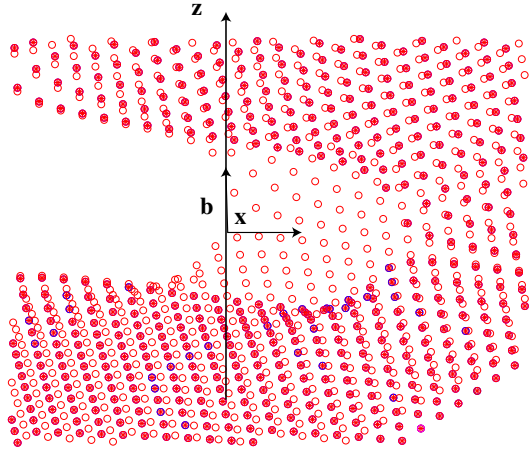}
\caption{(Color online) Superimposed zooms of the moving core of a screw dislocation.}
\label{screw3}
\end{center}\end{figure}

The continuum displacement field of a dislocation, ${\bf \tilde{u}} =(\tilde{u}_{1},
\tilde{u}_{2},\tilde{u}_{3})$, can be calculated as a stationary solution of the 
anisotropic Navier equations with a singularity $\propto r^{-1}$ at the dislocation core and 
such that $\int_{\mathcal{C}} (d{\bf x}\cdot \nabla) {\bf \tilde{u}} = - {\bf b}$, 
where ${\bf b}$ is the Burgers  vector and $\mathcal{C}$ is any closed curve encircling the
dislocation line \cite{ll7}. A pure screw dislocation along the $z$ axis with Burgers vector 
${\bf b}= (0,0,b)$ has a displacement vector ${\bf \tilde{u}} =(0,0,\tilde{u}_{3}(x,
y))$ \cite{nab67}. Then the strain energy density (\ref{e1}) becomes $W= C_{44}\,
|\nabla \tilde{u}_{3}|^2/2$, and the stationary equation of motion is $\Delta \tilde{u
}_{3} =0$. Its solution corresponding to a screw dislocation is $\tilde{u}_{3}(x,y)= b\, 
(2\pi)^{-1}\tan^{-1} (y/x)$ \cite{nab67}. The same symmetry considerations for Eq.\
(\ref{e13}) yield the following discrete equation for the $z$ component of the {\em 
nondimensional} displacement $u_{3}(l,m;t)$:
\begin{eqnarray}
M\, \ddot{u_{3}} &=& C_{44}\, \{D^-_1 [g(D^+_1 u_{3})\, g'(D^+_1 u_{3})] 
+ D^-_2 [g(D^+_2 u_{3})\, g'(D^+_2 u_{3})]\}. \label{dy1}
\end{eqnarray}
Numerical solutions of Eq.\ (\ref{dy1}) show that a static screw dislocation moves if 
an applied shear stress surpasses the static Peierls stress, $|F|>F_{cs}$, but that a moving 
dislocations continues doing so until the applied shear stress falls below a lower threshold 
$F_{cd}$ (dynamic Peierls stress); see Ref.~\cite{car03} for a similar situation 
for edge dislocations. To find the static solution of this equation corresponding to a screw 
dislocation, we could minimize an energy functional. However, it is more efficient to solve 
the following overdamped equation:
\begin{eqnarray}
\beta\, \dot{u_{3}} &=& C_{44}\, \{D^-_1 [g(D^+_1 u_{3})\, g'(D^+_1 u_{3})] 
+ D^-_2 [g(D^+_2 u_{3})\, g'(D^+_2 u_{3})]\}.   \label{dy2}
\end{eqnarray}
The stationary solutions of Eqs.\ (\ref{dy1}) and (\ref{dy2}) are the same, but the 
solutions of (\ref{dy2}) relax rapidly to the stationary solutions if we choose appropriately 
the damping coefficient $\beta$. We solve Eq.\ (\ref{dy2}) with initial condition 
$u_{3}(l,m;0)= U_{3}(l,m)\equiv b\, (2\pi a)^{-1} \tan^{-1}(m/l)$, and with boundary 
conditions $u_{3}(l,m;t)= U_{3}(l,m)+ F\, m$ at the upper and lower boundaries of our 
lattice. At the lateral boundaries, we use zero-flux Neumann boundary conditions. Here 
$F$ is an applied dimensionless stress with $|F|<F_{cs}$; to obtain the 
dimensional stress we should multiply $F$ by $C_{44}$. For such small stress, the solution of 
Eq.\ (\ref{dy2}) relaxes to a static screw dislocation $u_{3}(l,m)$ with the desired far 
field. If $F=0$, Figure \ref{screw1} shows the helical structure adopted by the deformed 
lattice $(l,m,n+u_{3}(l,m))$ for the piecewise linear $g(x)$ of Eq.\ (\ref{e9})
with $\alpha=0.24$. The numerical solution shows that moving a dislocation requires that 
we should have $g'(D_{j}^+u_{3})<0$ (with either $j=1$ or 2) in (\ref{dy1}) or 
(\ref{dy2}) at its core \cite{car03}. This is harder to achieve as $\alpha$ decreases.  

The motion of a pure screw dislocation is somewhat special because its Burgers vector is 
parallel to the dislocation line. Any plane containing the Burgers vector can be a glide 
plane. Under a shear stress $F>F_{cs}$ directed along the $y$ direction, a screw dislocation 
moves on the glide plane $xz$. A moving screw dislocation has the structure of a discrete 
traveling wave in the direction $x$, with far field $u_{3}(l-ct,m)+Fm$; $c=c(F)$ is the 
dislocation speed; see Figure \ref{screw3}. This is similar to the case of edge dislocations 
in the simple IAC model \cite{kkl}, as discussed in \cite{car03} where the details of the 
analysis can be looked up.

\subsection{Edge dislocations}
\label{sec:edge}
To analyze edge dislocations in the simplest case, we consider a isotropic cubic crystal 
($C_{44}= (C_{11}- C_{12})/2$) with planar discrete symmetry, so that ${\bf u}(l,m;t)=
(u_{1}(l,m;t),u_{2}(l,m;t),0)$ is independent of $z=na$. 

To find the stationary edge dislocation of the discrete model, we first have to write the 
corresponding stationary edge dislocation of isotropic elasticity. An edge dislocation 
directed along the $z$ axis (dislocation line), and having Burgers vector $(b,0,0)$ has a 
displacement vector ${\bf \tilde{u}}= (\tilde{u}_{1}(x,y),\tilde{u}_{2}(x,y),0)$ 
with a singularity $\propto r^{-1}$ at the core and satisfying $\int_{\mathcal{ C}} 
(d{\bf x}\cdot\nabla) {\bf \tilde{u}} =- (b,0,0)$, for any closed curve $\mathcal{C}$ 
encircling the $z$ axis. It satisfies the planar stationary Navier equations with a singular 
source term:
\begin{eqnarray}
\Delta {\bf \tilde{u}} + {1\over 1-2\nu} \nabla(\nabla\cdot {\bf \tilde{u}}) = -
(0,b,0)\,\delta(r).    \label{dy3}
\end{eqnarray}
Here $r=\sqrt{x^2+y^2}$ and $\nu=\lambda/[2(\lambda + \mu)]$ is the Poisson ratio; 
cf.\ page 114 of Ref.~\cite{ll7}. The appropriate solution is 
\begin{eqnarray} 
\tilde{u}_{1} &= & {b \over 2 \pi} \left[\tan^{-1}\left({y\over x}\right) + 
{ xy \over 2(1-\nu)(x^2+y^2)}\right],\nonumber\\  
\tilde{u}_{2} &= & {b \over 2 \pi} \left[ -{1-2\nu\over 4(1-\nu)}\,\ln\left( 
{x^2+y^2\over b^2}\right) + {y^2 \over 2(1-\nu) (x^2+y^2)}\right], \label{dy4}
\end{eqnarray}
cf.\ Ref.~\cite{nab67}, pag.\ 57. 

Eqs.\ (\ref{dy4}) yield the nondimensional static displacement vector ${\bf U}(l,m)= 
(\tilde{u}_{1}(la,ma)/a,\tilde{u}_{2}(la,ma)/a,0)$, which will be used to find the 
stationary edge dislocation of the discrete equations of motion. For this planar configuration, 
the conservative equations of motion (\ref{e13})  become
\begin{eqnarray} 
M\, \ddot{u}_{1} &=& C_{11}\, D^-_1 [g(D^+_1 u_{1})\, g'(D^+_1 u_{1})] +
C_{12}\, D^-_1 [g(D^+_2 u_{2})\, g'(D^+_1 u_{1})] \nonumber\\
& &+ C_{44}\, D^-_2 \{[g(D^+_2 u_{1})+ g(D^+_1 u_{2})]\, g'(D^+_2 u_{1})\},
\label{dy5}\\   
M\, \ddot{u}_{2} &=& C_{11}\, D^-_2 [g(D^+_2 u_{2})\, g'(D^+_2 u_{2})] +
C_{12}\, D^-_2 [g(D^+_1 u_{1})\, g'(D^+_2 u_{2})]\nonumber\\ 
&& + C_{44}\, D^-_1 \{[g(D^+_1 u_{2}) + g(D^+_2 u_{1})]\, g'(D^+_1 u_{2})\}.
\label{dy6}
\end{eqnarray}  

To find the stationary edge dislocation corresponding to these equations, we set $C_{44}= 
(C_{11}- C_{12})/2$ (isotropic case), and replace the inertial terms $M\ddot{u}_{1}$ and 
$M\ddot{u}_{2}$ by $\beta\dot{u}_{1}$ and $\beta\dot{u}_{2}$, respectively. The 
resulting overdamped equations,
\begin{eqnarray} 
\beta\, \dot{u}_{1} &=& C_{11}\, D^-_1 [g(D^+_1 u_{1})\, g'(D^+_1 u_{1})] +
C_{12}\, D^-_1 [g(D^+_2 u_{2})\, g'(D^+_1 u_{1})] \nonumber\\
& &+ C_{44}\, D^-_2 \{[g(D^+_2 u_{1})+ g(D^+_1 u_{2})]\, g'(D^+_2 u_{1})\},
\label{dy7}\\   
\beta\, \dot{u}_{2} &=& C_{11}\, D^-_2 [g(D^+_2 u_{2})\, g'(D^+_2 u_{2})] +
C_{12}\, D^-_2 [g(D^+_1 u_{1})\, g'(D^+_2 u_{2})]\nonumber\\ 
&& + C_{44}\, D^-_1 \{[g(D^+_1 u_{2}) + g(D^+_2 u_{1})]\, g'(D^+_1 u_{2})\}, 
\label{dy8}
\end{eqnarray}  
have the same stationary solutions as Eqs.\ (\ref{dy5}) and (\ref{dy6}). We solve Eqs.\ 
(\ref{dy7}) and (\ref{dy8}) with initial condition ${\bf u}(l,m;0) = {\bf U}(l,m)$ given 
by Eqs.\ (\ref{dy4}), and with boundary conditions ${\bf u}(l,m;t) = {\bf U}(l,m)+ (F\, 
m,0,0)$ at the upper and lower boundaries of the lattice ($F$ is a dimensionless applied shear 
stress; recall that the displacement vector in the discrete equations is always dimensionless). 
If $|F|<F_{cs}$ ($F_{cs}$ is the static Peierls stress for edge dislocations), the 
solution of Eqs.\ (\ref{dy7}) and (\ref{dy8}) relaxes to a static edge dislocation 
$(u_{1}(l,m),u_{2}(l,m),0)$ with the appropriate continuum far field.

 \begin{figure}
\begin{center}
\includegraphics[width=5cm]{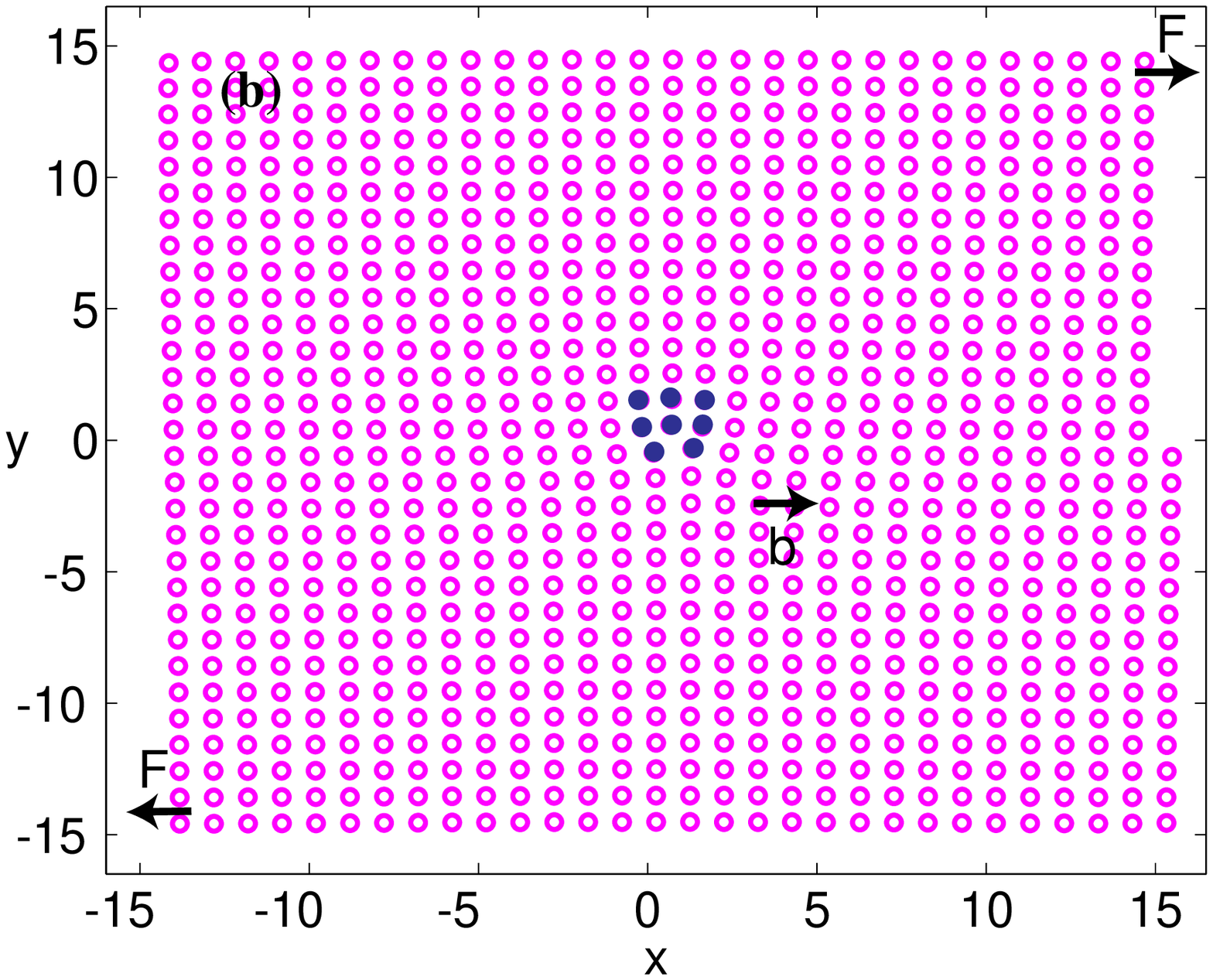}%{aris17.eps}
\includegraphics[width=5cm]{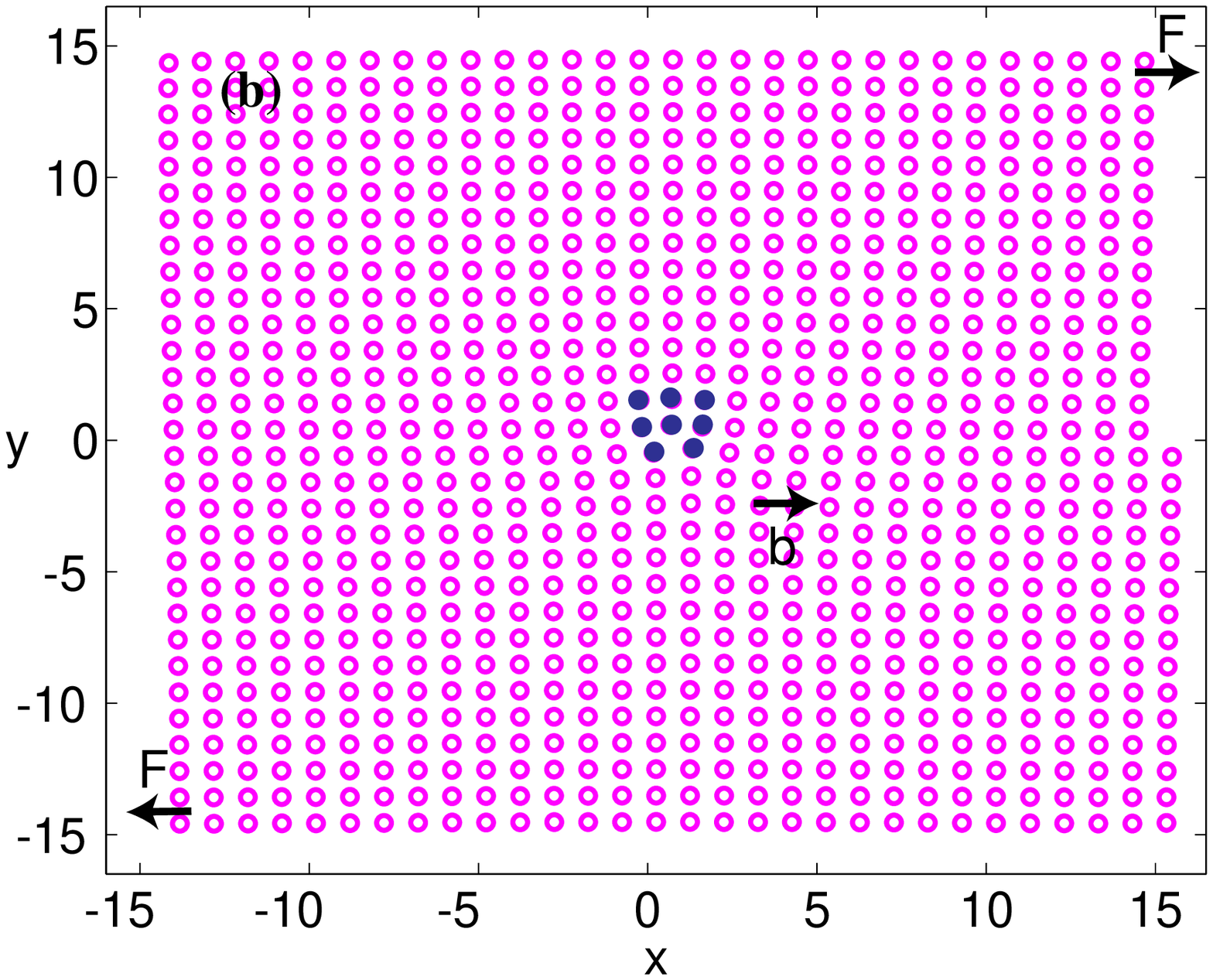}%{aris18.eps}
\includegraphics[width=5cm]{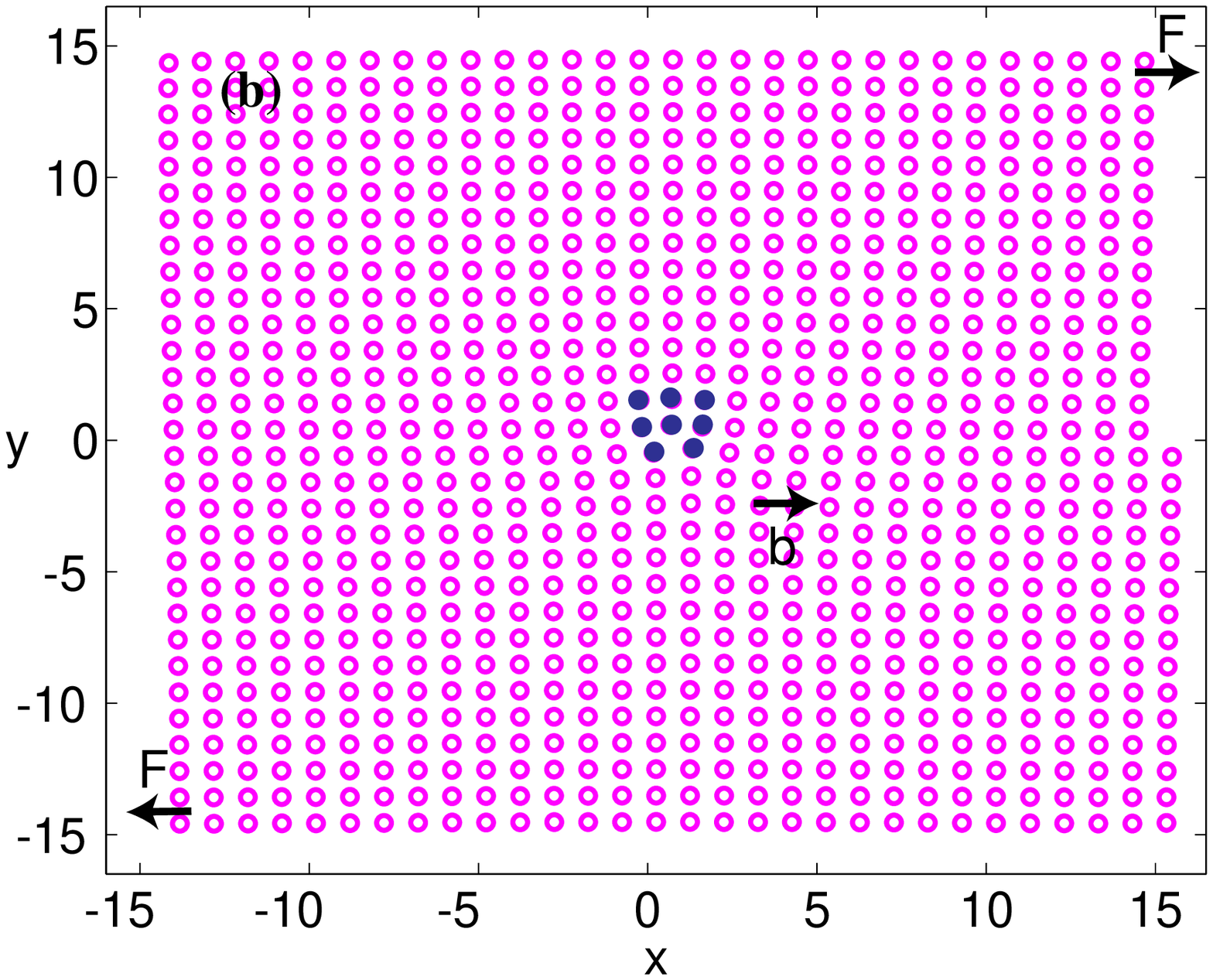}%{aris20.eps}
\end{center}
\caption{(Color online) Edge dislocation for the piecewise linear $g(x)$ of Eq.\ 
(\ref{e9}) with (a) $\alpha=0.27$, (b) $\alpha=0.29$, (c) $\alpha=0.32$.}
\label{fig:edge31}
\end{figure}
 \begin{figure}
\begin{center}
\includegraphics[width=8cm]{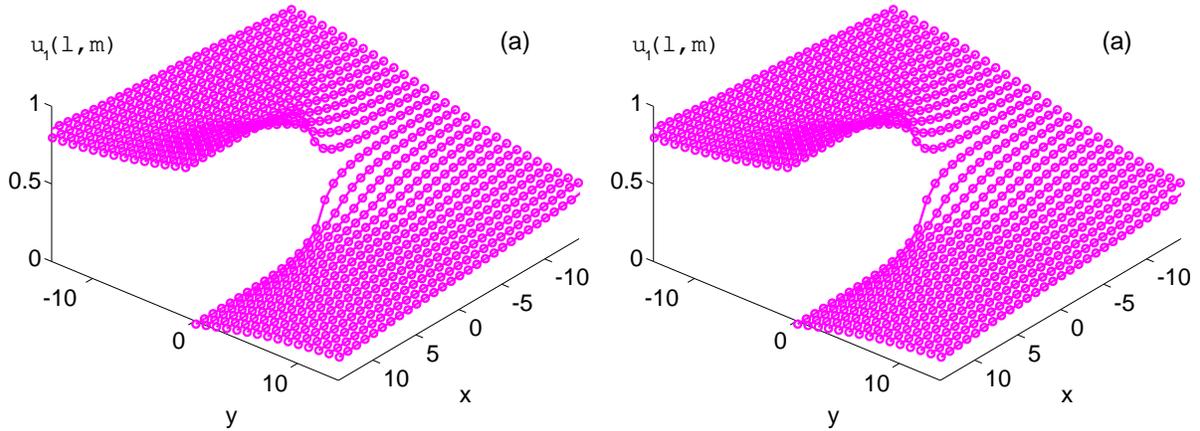}%{uaris16.eps}
\includegraphics[width=8cm]{fig5a.eps}%{uaris20.eps}
\end{center}
\caption{(Color online) Profile of $u_{1}(l,m)$ for an edge dislocation with the piecewise 
linear $g(x)$ of Eq.\ (\ref{e9}) when  (a) $\alpha=0.24$,  (b) $\alpha=0.32$.}
\label{fig:edge32}
\end{figure}
\begin{figure}
\begin{center}
\includegraphics[width=4cm]{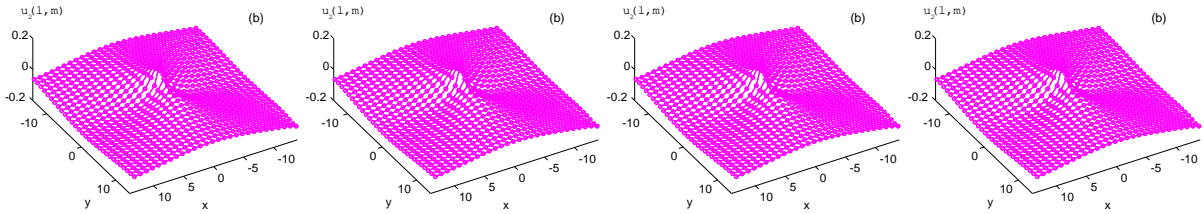}%{varis16.eps}
\includegraphics[width=4cm]{fig15a.eps}%{varis17.eps}
\includegraphics[width=4cm]{fig15a.eps}%{varis18.eps}
\includegraphics[width=4cm]{fig15a.eps}%{varis20.eps}
\end{center}
\caption{(Color online) Profile of $u_{2}(l,m)$ for an edge dislocation with the piecewise 
linear $g(x)$ of Eq.\ (\ref{e9}) when  (a) $\alpha=0.24$, (b) $\alpha=0.27$,
 (c) $\alpha=0.29$, (d) $\alpha=0.32$.}
\label{fig:edge33}
\end{figure}

In our numerical calculations of the static edge dislocation, we use the elastic constants 
of tungsten (which is an isotropic bcc crystal), $C_{11}=521$ GPa, $C_{12}=201$ GPa, 
$C_{44}=160$ GPa ($C_{11}= C_{12}+ 2C_{44}$) \cite{hir82}. This yields $\nu= 
0.278$.
Figure \ref{fig:edge31} shows the structure adopted by the deformed lattice 
$(l+u_{1}(l,m),m+u_{2}(l,m))$ when $\nu= 0.278$. for the asymmetric piecewise linear 
function $g(x)$ with three different values of $\alpha$. The profiles of the displacement 
vector are shown in Figures \ref{fig:edge32} and \ref{fig:edge33}. The glide motion of 
edge dislocations occurs on the glide plane defined by their Burgers vector and the 
dislocation line, and in the direction of the Burgers vector. In our case, a shear stress in the 
direction $y$, will move the dislocation in the direction $x$. For conservative or damped 
dynamics, the applied shear stress has to surpass the static Peierls stress to depin a static 
dislocation, and a moving dislocation propagates provided the applied stress is larger than 
the dynamic Peierls stress (smaller than the static one) \cite{car03}. Numerical solutions 
show that the dimensionless Peierls stress depends on $\alpha$ in Eq.\ (\ref{e9}) as shown 
in Fig.~\ref{fcritico}. As the interval of $x$ for which $g'(x)<0$ shrinks (which occurs as 
$\alpha$ decreases), the static Peierls stress increases and the dislocation becomes harder to 
move. The size of the dislocation core is also related to the shape of $g(x)$. As shown by 
Figures \ref{fig:edge31} to \ref{fig:edge33}, the dislocation core expands as $\alpha$ 
increases: for $\alpha\leq 0.26$, the dislocation core is very narrow, as shown in 
Figs.~\ref{fig:edge31}(a) and \ref{fig:edge33}(a). Figs.~\ref{fig:edge31}(b), and
\ref{fig:edge33}(a) to \ref{fig:edge33}(c) show that the dislocation core widens one 
lattice point as $\alpha$ sweep the interval $0.26<\alpha<0.29$, in which the variation of 
the Peierls stress is very small (see the {\em plateau} in Fig.~\ref{fcritico}). The dislocation 
core gains more lattice points as $\alpha$ increases to 0.32 and beyond, cf.\ 
Figs.~\ref{fig:edge31}(c),
\ref{fig:edge32}(b) and \ref{fig:edge33}(d). Thus the size of the dislocation core and the 
Peierls stress are related to the width of the interval of $x$ for which $g'(x)<0$ and to the 
actual value of the slope. As an additional example, the symmetric sine function $g(x)=\sin
(2\pi x)/(2\pi)$ has wide subintervals of small slope, which produces very low Peierls 
stresses and an artificially wide dislocation core. In \cite{kkl,car03}, this problem was 
avoided by setting $u_{2}=0$ and changing $g(D^+_1 u_{1})\leftrightarrow D^+_1 
u_{1}$ and $2\pi g(D^+_2 u_{1}) = \sin (2\pi D^+_1 u_{1})$ in (\ref{dy5}) and 
(\ref{dy7}). A moving dislocation is a discrete traveling wave advancing along the $x$ axis, 
and having far field $(u_{1}(l-ct,m)+Fm,u_{2}(l-ct,m))$. The analysis of depinning and 
motion of planar edge dislocations follows that explained in Ref.~\cite{car03} with 
technical complications due to our more complex discrete model. 
 
 \subsection{Interaction of edge dislocations and crack formation}
\label{sec:cracks}
\begin{figure}\begin{center}
\includegraphics[width=8cm]{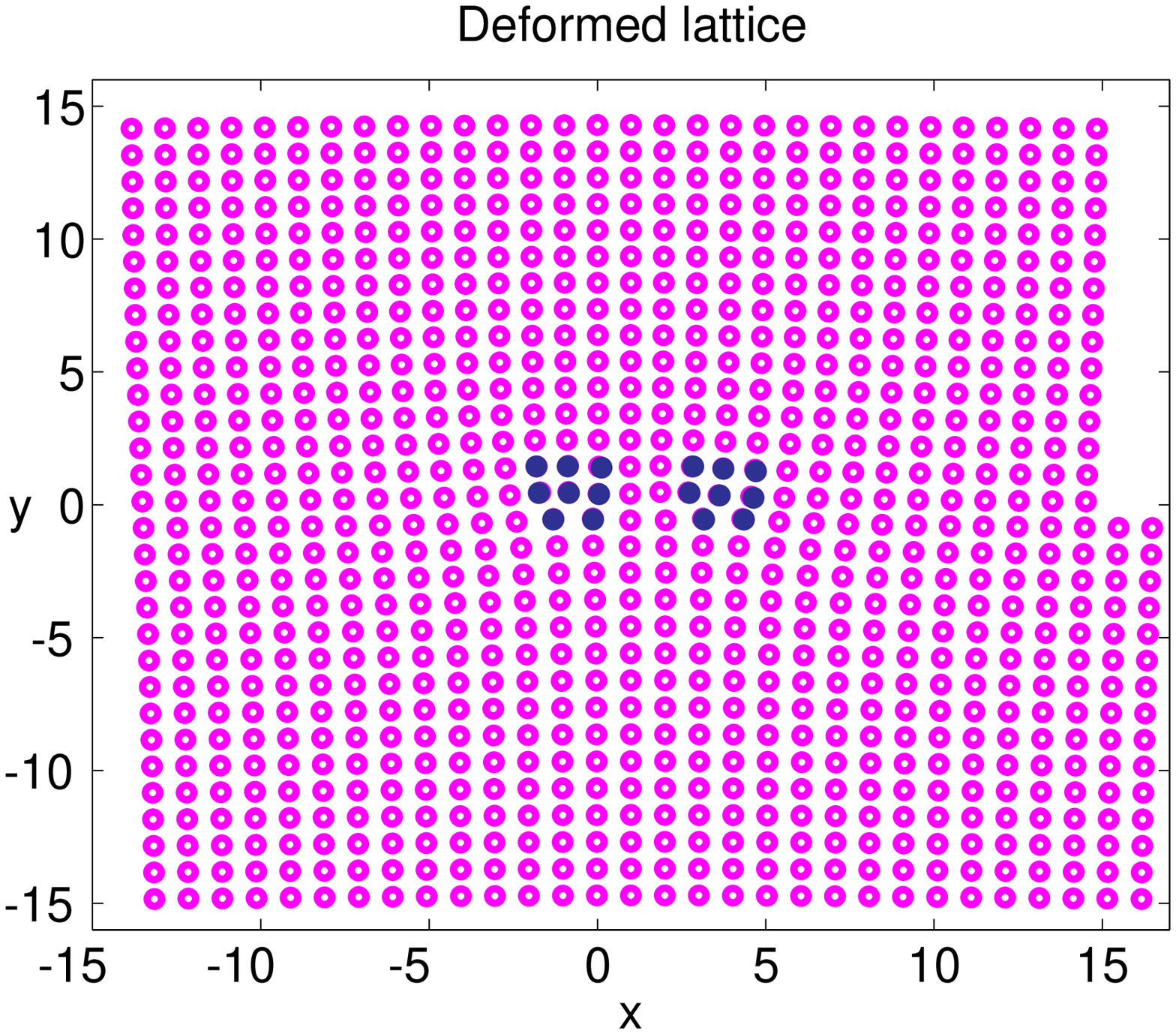}
\includegraphics[width=8cm]{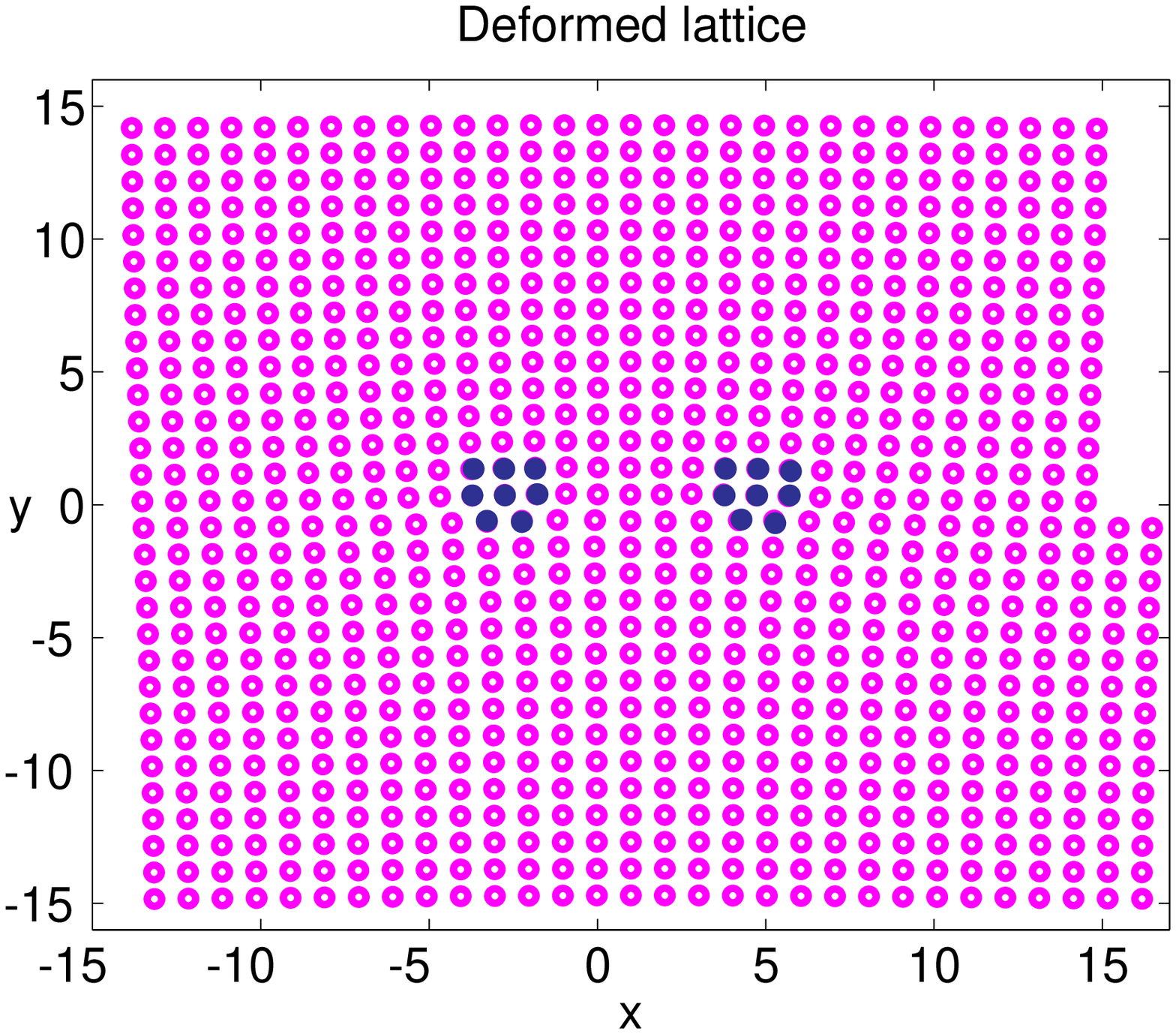}
\end{center}
\caption{(Color online) Repulsion of like sign edge dislocations.}
\label{repulsion}
\end{figure}

\begin{figure}\begin{center}
\includegraphics[width=8cm]{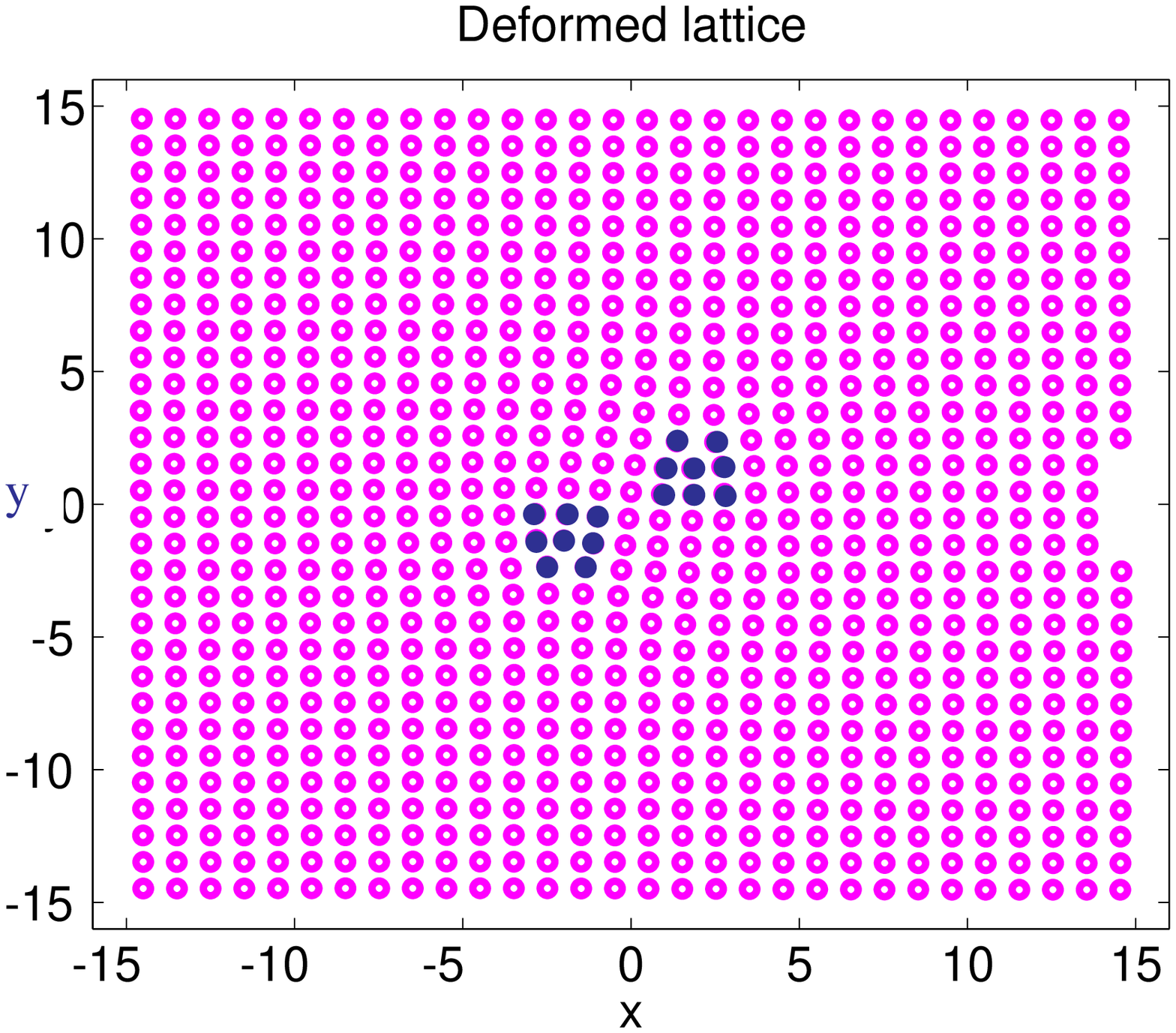}
\includegraphics[width=8cm]{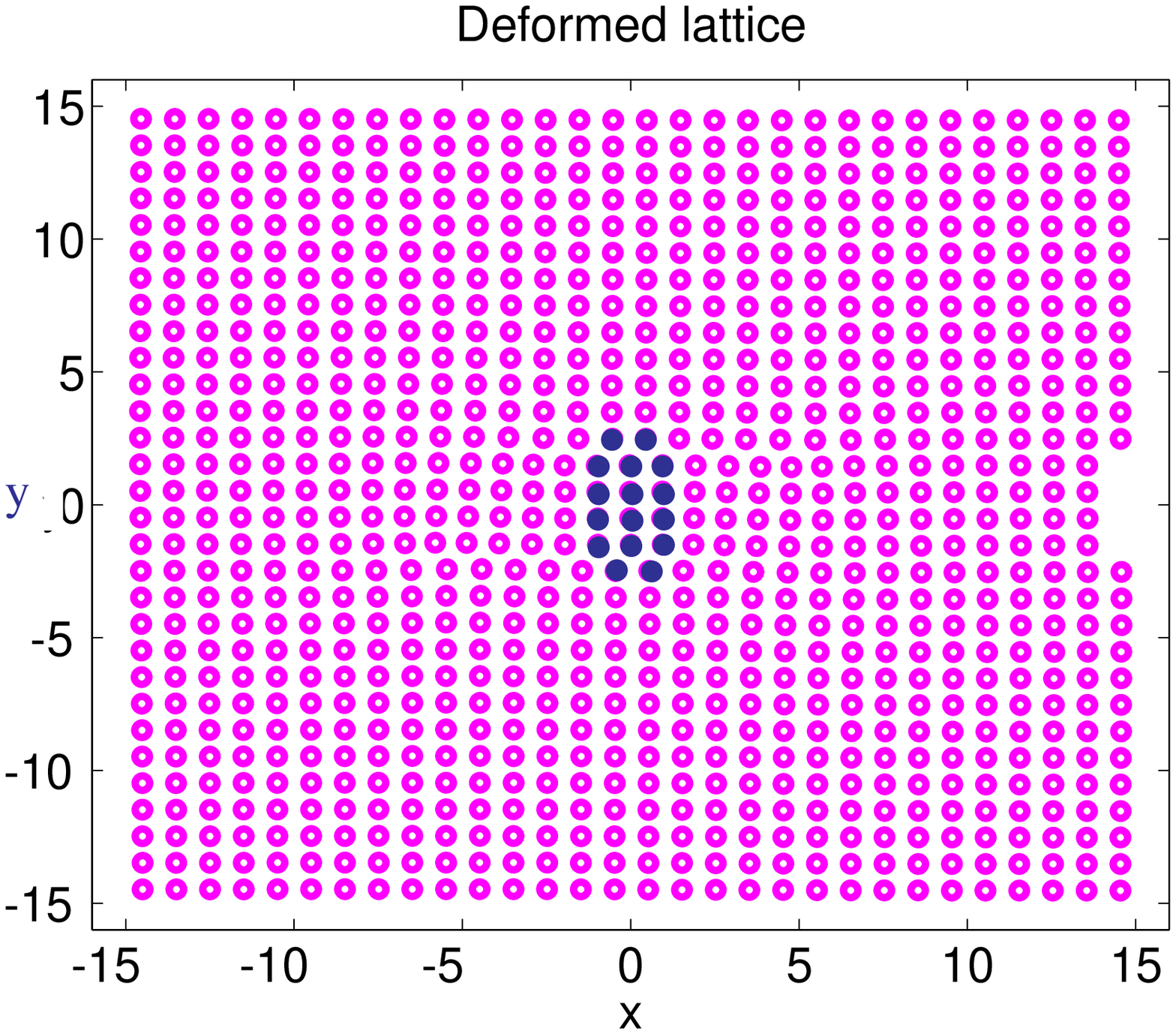}
\end{center}
\caption{(Color online) Atraction of opposite sign edge dislocations
leading to formation of a dislocation loop.}
\label{loop}
\end{figure}

\begin{figure}\begin{center}
\includegraphics[width=8cm]{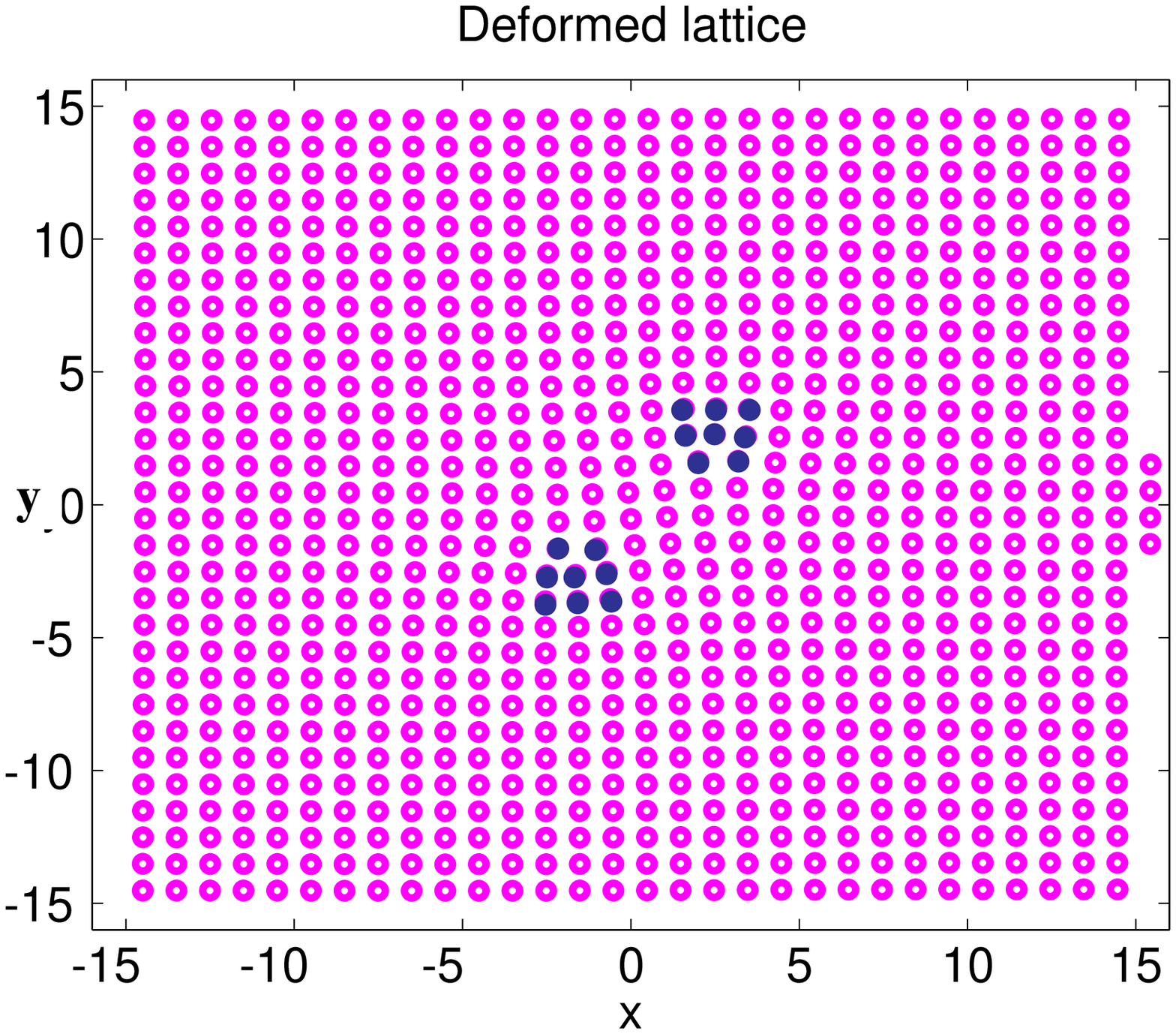}
\includegraphics[width=8cm]{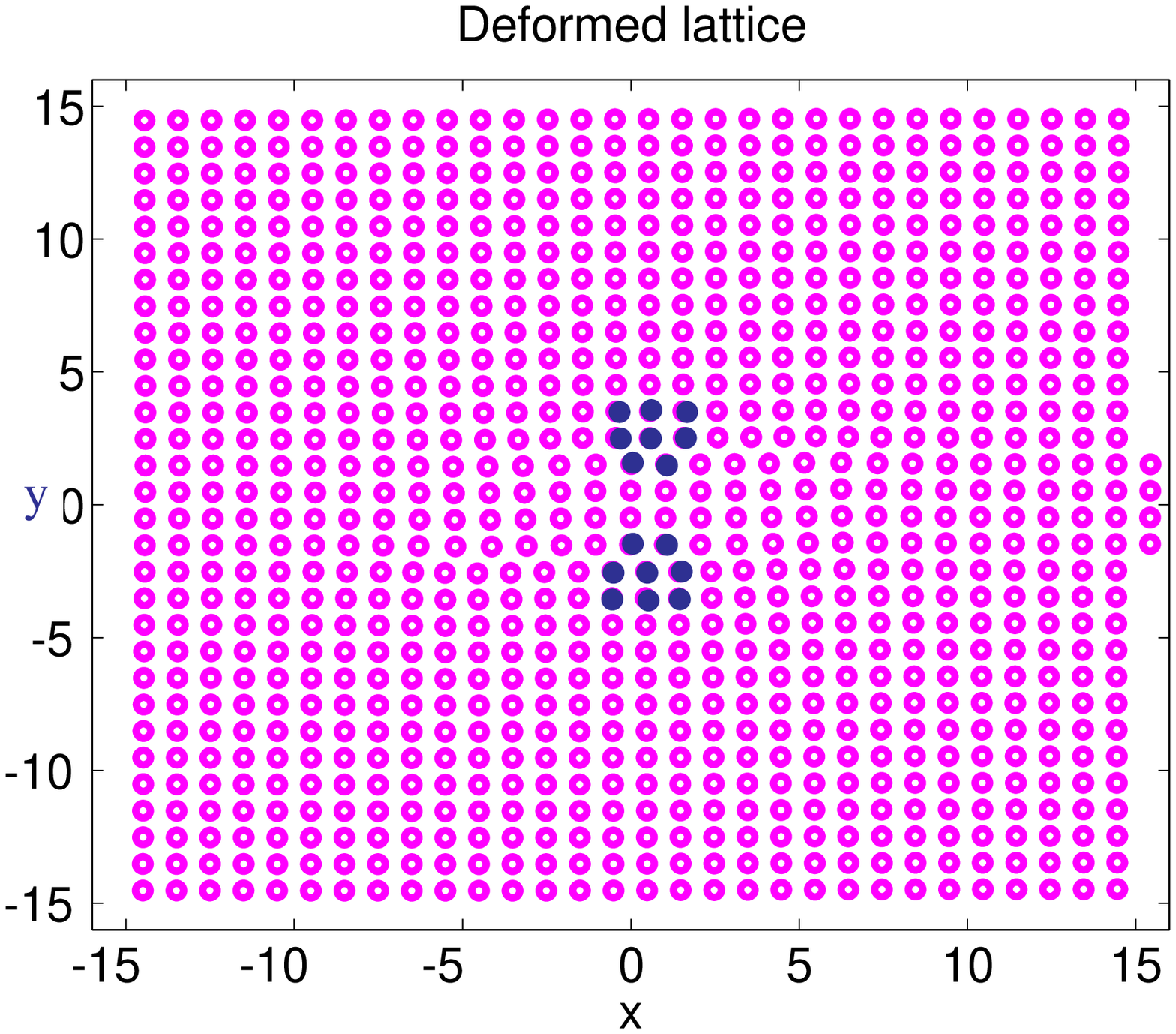}
\end{center}
\caption{(Color online) Atraction of opposite sign edge dislocations
leading to formation of a dislocation dipole.}
\label{dipole}
\end{figure}

The reduced 2D model (\ref{dy5}) - (\ref{dy6}) can be solved numerically to
illustrate interaction of edge dislocations. Figure \ref{repulsion}
illustrates the repulsion of equal-sign edge dislocations whereas opposite
edge dislocations attract each other and form dislocation loops as in Fig. 
\ref{loop} or dislocation dipoles as in Fig. \ref{dipole}. Friction terms
affect numerical simulations of the model as follows. As in the case of 1D
models \cite{wavy}, atoms may oscillate far from the core of a moving
dislocation when the equations of motion are conservative or slightly damped. 
Large friction (order one coefficients) reduces the oscillations of individual
atoms, the instantaneous position of the core of the defect is easier to
locate and its  movement in the distorted lattice is easier to follow. Small
friction (order $10^{-2}$ coefficients) results in dislocation glide combined
with oscillations of the individual atoms. The figures \ref{repulsion}, \ref{loop}
and \ref{dipole} were obtained with small friction. See Refs.~\onlinecite{wavy} and 
\onlinecite{carPRL01} on the impact of friction and inertia on 1D wave front profiles.


\begin{figure}\begin{center}
\includegraphics[width=8cm]{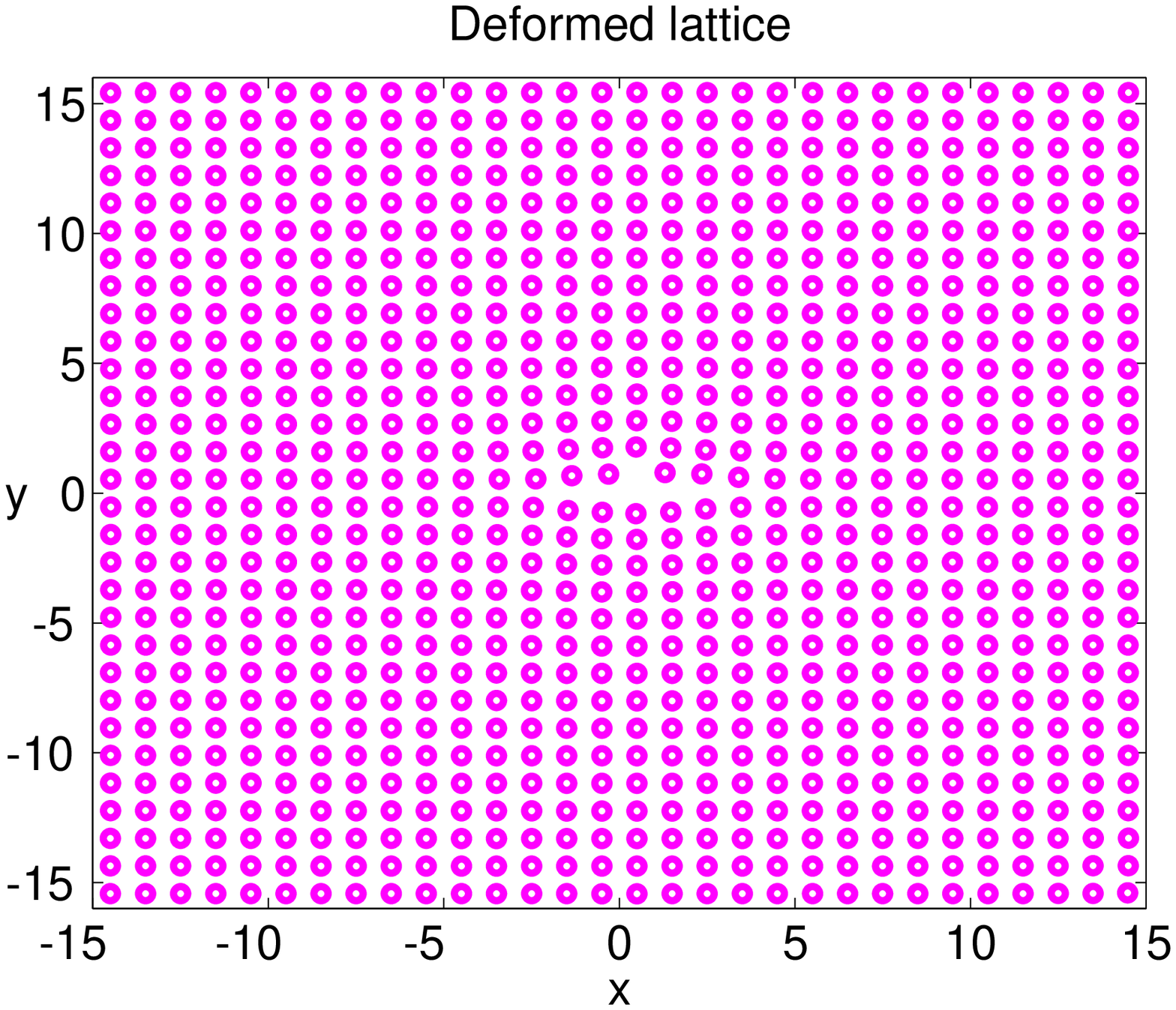}
\includegraphics[width=8cm]{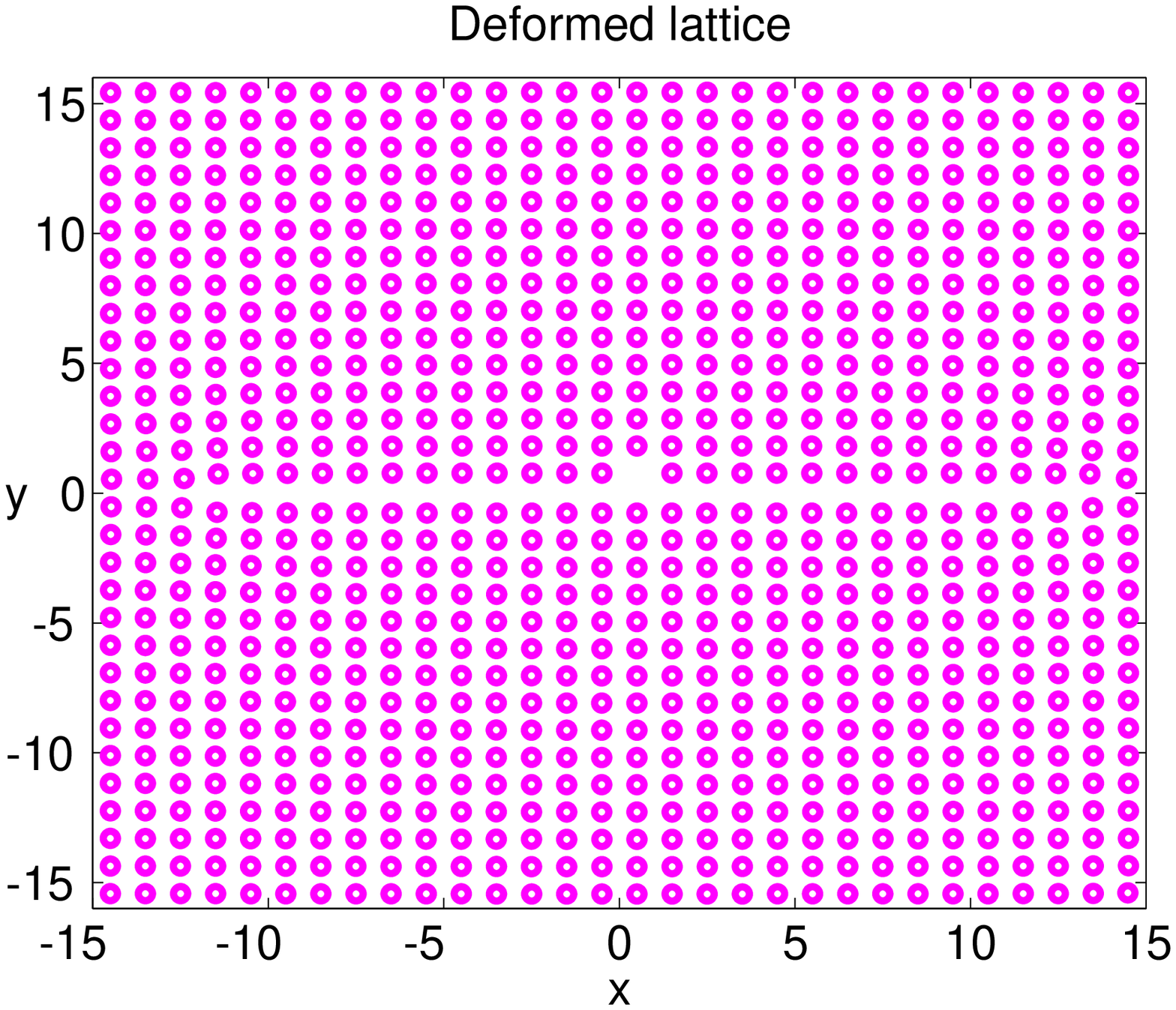}
\end{center}
\caption{(Color online) Snapshots showing crack generation and growth induced by 
applying a tension in the direction $y$ to a dislocation dipole.}
\label{crack}
\end{figure}

Figure \ref{crack} shows the formation of a crack propagating in the $x$
direction under an applied tension in the $y$ direction. In principle,
numerically solving the discrete equations of motion we can find:  (i)
the threshold stress for crack propagation, (ii) the direction of propagation,
(iii) the crack speed, and (iv) the crack shape. We do not need to impose
additional conditions such as displacement thresholds for breaking atomic
bonds as in the usual spring models for brittle fracture \cite{sle81,guinea}.

\section{Elasticity in a non-orthogonal basis}
\label{sec:fcc}
\subsection{Equations of motion}
For  fcc or bcc crystals, the primitive vectors of the unit cell are not orthogonal. To find a 
discrete model for these crystals, we should start by writing the strain energy density in a 
non-orthogonal vector basis, $a_{1}$, $a_{2}$, $a_{3}$, in terms of the usual orthonormal 
vector basis $e_1$, $e_2$, $e_3$ determined by the cube sides of length $a$. Let $x_i$ denote 
coordinates in the basis $e_i$, and let $x'_i$ denote coordinates in the basis $a_i$. Notice that 
the $x_i$ have dimensions of length while the $x'_i$ are dimensionless. The matrix $T=(a_1,
a_2,a_3)$ whose columns are the coordinates of the new basis vectors in terms of the old 
orthonormal basis can be used to change coordinates as follows:
\begin{eqnarray}
x_i'=T_{ij}^{-1}x_j, \; x_i=T_{ij}x'_j . \label{fcc2}
\end{eqnarray}
Similarly, the displacement vectors in both basis are related by
\begin{eqnarray}
u_i'=T_{ij}^{-1}u_j, \; u_i=T_{ij}u'_j , \label{fcc3}
\end{eqnarray}
and partial derivatives obey
\begin{eqnarray}
{\partial \over \partial x_i'}=T_{ji} {\partial \over \partial x_j},
 \;{\partial \over \partial x_i}=T_{ji}^{-1} {\partial \over \partial x_j'} .
\label{fcc4}
\end{eqnarray}
By using these equations, the strain energy density $W= (1/2) c_{iklm} e_{ik} e_{lm}$ 
can be written as
\begin{eqnarray}
W= {1\over 2}\, c_{ijlm}{\partial u_i \over \partial x_j}{\partial u_l \over
\partial x_m}= {1\over 2}\, c_{rspq}' {\partial u_r' \over \partial x_s'}
{\partial u_p' \over \partial x_q'}, \label{fcc5}
\end{eqnarray}
where the new elastic constants are:
\begin{eqnarray}
c_{rspq}'=c_{ijlm}T_{ir}T_{sj}^{-1}T_{lp}T_{qm}^{-1}. \label{fcc6}
\end{eqnarray}
Notice that the elastic constants have the same dimensions in both the orthogonal and the 
non-orthogonal basis. To obtain a discrete model, we shall consider that the dimensionless
displacement vector $u'_{i}$ depends on dimensionless coordinates $x'_{i}$ that are integer
numbers $u'_{i}= u'_{i}(l,m,n;t)$. As in the case of sc crystals, we replace the distortion 
tensor (gradient of the displacement vector in the non-orthogonal basis) by a periodic 
function of the corresponding forward difference, $w_{i}^{(j)} = g(D^+_j u'_i)$, cf.\ 
Eq.\ (\ref{e9}). The discretized strain energy density is
\begin{eqnarray} 
W(l,m,n;t) = {1\over 2 }\, c_{rspq}' g(D^+_s u_r')\, g(D^+_q u_p'). \label{fcc7}
\end{eqnarray}
The elastic constants $c_{rspq}'$ can be calculated in terms of the Voigt stiffness constants 
for a cubic crystal, $C_{11}$, $C_{44}$ and $C_{12}$. Eq.\ (\ref{e2}) yields
$c_{ijlm}= C_{44}(\delta_{il}\delta_{jm}+\delta_{im}\delta_{lj})
+ C_{12}\delta_{ij}\delta_{lm}-H (\delta_{1i}\delta_{1j}\delta_{1l}\delta_{1m}
+\delta_{2i}\delta_{2j}\delta_{2l}\delta_{2m} + \delta_{3i}
\delta_{3j}\delta_{3l}\delta_{3m})$,
where $H=2C_{44}+C_{12}-C_{11}$ measures the anisotropy of the crystal and Eq.\ 
(\ref{fcc6}) provides the tensor $c_{rspq}'$. The elastic energy can be obtained from Eq.\
(\ref{fcc7}) for $W$ by means of Eqs.\ (\ref{e8}). Then the equations of motion 
(\ref{e10}) are
\begin{eqnarray}
\rho a^3\, {\partial^2 u'_i \over \partial t^2}= -T^{-1}_{iq}T_{pq}^{-1} 
{\partial V \over \partial u_p'}, \nonumber 
\end{eqnarray}
which, together with Eqs.\ (\ref{e8}) and (\ref{fcc7}), yield
\begin{eqnarray}
\rho\, {\partial^2 u'_i \over \partial t^2}= T^{-1}_{iq}T_{pq}^{-1}\,
D^-_j [c_{pjrs}'\, g'(D^+_ju_p')\, g(D^+_s u_r')] . \label{fcc8}
\end{eqnarray}
This equation becomes (\ref{e13}) for orthogonal coordinates, $T^{-1}_{iq}=
\delta_{iq}/a$, once we take into account the Einstein convention on summation over
repeated indices in (\ref{e13}).

\subsection{Far field of a dislocation}
As in the case of sc crystals studied in Section \ref{sec:dynamics}, we should determine the 
elastic far field of a dislocation under zero stress to set up the initial and boundary data 
needed to solve numerically the discrete equations of motion (\ref{fcc8}). We can calculate 
the elastic far field of any straight dislocation following the method explained in Chapter 13 
of Hirth and Lothe's book \cite{hir82}. Firstly, we determine the elastic constants in an 
orthonormal coordinate system $e_1''$, $e_2''$, $e_3''$ with $e_3'' = -\xi$ parallel to the 
dislocation line. The result is
\begin{eqnarray}
c_{ijkl}''=c_{ijkl} - H \sum_{n=1}^3 (S_{in}S_{jn}S_{kn}S_{ln}
-\delta_{in}\delta_{jn}\delta_{kn}\delta_{ln}). \label{fcc9}
\end{eqnarray}
Here the rows of the orthogonal matrix $S= (e_1'',e_2'',e_3'')^t$ are the coordinates 
of the $e_i''$'s in the old orthonormal basis $e_1$, $e_2$, $e_3$. In these new coordinates, 
the elastic displacement field $(u_1'',u_2'',u_3'')$ depends only on $x_1''$ and on $x_2''$. 
The Burgers vector and the elastic displacement field satisfy $b''_1=b''_2=0$ and 
$u_1''=u_2''=0$ for a pure screw dislocation in an infinite medium. For a pure edge
dislocation, $b''_3=0$ and $u_3''=0$. 
Secondly, the displacement vector $(u_1'',u_2'',u_3'')$ is calculated as follows:
\begin{itemize}
\item Select three roots $p_1,p_2,p_3$ with positive imaginary
part out of each pair of complex conjugate roots of the polynomial
$det[a_{ik}(p)]=0$, $a_{ik}(p)=c_{i1k1}''+(c_{i1k2}''+c_{i2k1}'')p+
c''_{i2k2}p^2$.
\item For each $n=1,2,3$ find an eigenvector $A_k(n)$ associated to the
zero eigenvalue for the matrix $a_{ik}(p_n)$.
\item Solve Re$\sum_{n=1}^3 A_k(n)D(n)=b''_k,$ $k=1,2,3$ and
Re$\sum_{n=1}^3 \sum_{k=1}^3 (c''_{i2k1}+c''_{i2k2} p_n)A_k(n)D(n) = 0$,
$i=1,2,3$ for the imaginary and real parts of $D(1)$,$D(2)$,$D(3)$.
\item For $k=1,2,3$, $u_k''=$ Re$[-{1\over 2\pi i}\sum_{n=1}^3
A_k(n) D(n)\ln (x_1''+p_n x_2'')]$.
\end{itemize}
Lastly, we can calculate the displacement vector $u_k'$ in the non-orthogonal basis $a_i$ 
from $u_k''$.

\subsection{Discrete models for fcc metals}
\label{ssec:fcc}
For fcc metals, the non-orthogonal vector basis comprising primitive vectors is 
\begin{eqnarray}
a_1= {a\over 2}\, (1,1,0), \; a_2= {a\over 2}\, (0,1,1), \; a_3={a\over 2}\, (1,0,1).
\label{fcc1}
\end{eqnarray}
The equations of motion are (\ref{fcc8}) with the 
corresponding transformation matrix $T=(a_1,a_2,a_3)$.

We shall now analyze the motion of dislocations in the case of gold. The initial and 
boundary data for the numerical simulations are constructed from the far fields of 
dislocations in anisotropic elasticity as explained in the previous Subsection. We 
have considered two straight dislocations: the perfect edge dislocation directed 
along $\xi= (-1,1,-2)/\sqrt{6}$ (with a Burgers vector which is one of the translation 
vectors of the lattice, and therefore glide of the dislocation leaves behind a perfect crystal 
\cite{hul01}) and the pure screw dislocation along $\xi= (1,1,0)/\sqrt{2}$. For the 
perfect edge dislocation, we select:
\begin{eqnarray}
e_1''=(-1,-1,0)/\sqrt{2},\; e_2''=(1,-1,-1)/\sqrt{3}, \; e_3''= (1,-1,2)/\sqrt{6},
\label{fcc10}
\end{eqnarray}
which are unit vectors parallel to the Burgers vector ${\bf b}$, the normal to the glide 
plane ${\bf n}$, and minus the tangent to the dislocation line $-\xi$, respectively. 
For the pure screw dislocation, we have:
\begin{eqnarray}
e_1''=(1,-1,2)/\sqrt{6}, \; e_2''=(-1,1,1)/\sqrt{3},\; e_3''=(-1,-1,0)/\sqrt{2} ,
\label{fcc11}
\end{eqnarray}
where $e_2''$ is a unit vector normal to the glide plane and $e_3''$ is a unit vector parallel 
to the dislocation line and to the Burgers vector (but directed in the opposite sense). 

\begin{figure}\begin{center}
\includegraphics[width=8cm]{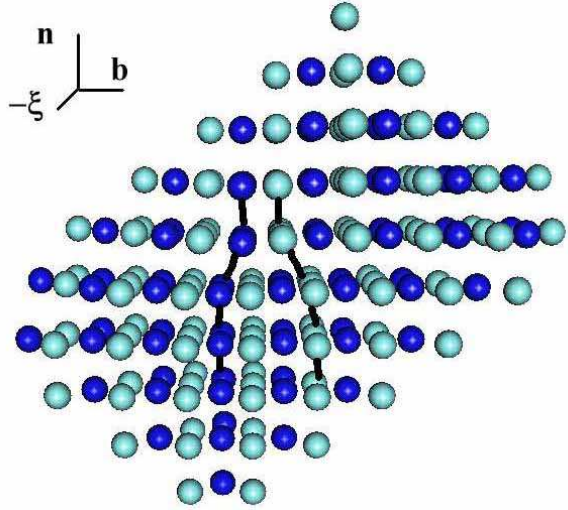}
\end{center}
\caption{(Color online) Perfect edge dislocation in a gold lattice displaying a two-fold 
stacking sequence of planes. Lines locating the dislocation core are a guide for the eye.}
\label{gold1}
\end{figure}

\begin{figure}\begin{center}
\includegraphics[width=8cm]{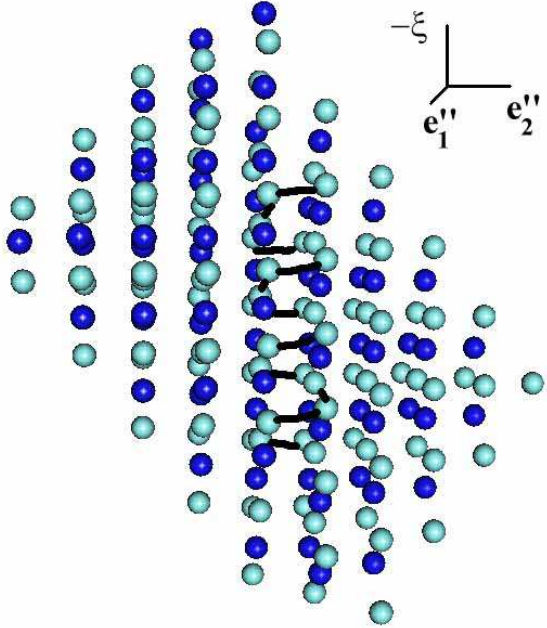}
\end{center}
\caption{(Color online) Screw dislocation in a gold lattice. Lines locating the dislocation 
core are a guide for the eye.}
\label{gold2}
\end{figure}

For gold, $C_{11}=186$ GPa, $C_{44}=42$ GPa, $C_{12}=157$ GPa and $H=55$ GPa.
The lattice constant is $a= 4.08$ \AA\, and
the density is $\rho=1.74 $ g/cm$^3$. Figures \ref{gold1} and \ref{gold2} show 
the perfect edge dislocation and the screw dislocation obtained as stationary solutions of 
model (\ref{fcc8}). Due to the boundary conditions we have chosen, their far fields 
match the corresponding elastic far fields of the dislocations (written in the non-orthogonal 
coordinates corresponding to the primitive cell vectors $a_1$, $a_2$, $a_3$). Dark and light 
colors are used to trace points placed in different planes in the original lattice. Note that the 
planes perpendicular 
to the Burgers vector in Fig.~\ref{gold1} have a two-fold stacking sequence 
`dark-light-dark-light \ldots' The extra half-plane of the edge dislocation consists of 
two half planes (one dark and one light) in the dark-light-dark-light \ldots\, sequence. 
Movement of this unit dislocation by glide retains continuity of the dark planes and the
light planes across the glide plane, except at the dislocation core where the extra half
planes terminate \cite{hul01}. 


\subsection{Discrete models for bcc metals}
\label{sec:bcc}

The discrete model for bcc metals is similar to that for fcc metals explained in the previous
Subsection, but the non-orthogonal vector basis comprising primitive vectors is now 
\begin{eqnarray}
a_1 = {a\over 2}\, (1,1,1), \; a_2= {a\over 2}\,(-1,1,1), \; a_3= {a\over 2}\,(1,-1,1).
\label{bcc1}
\end{eqnarray}
The equations of motion are (\ref{fcc8}) with the corresponding transformation 
matrix $T=(a_1,a_2,a_3)$.

As in the previous Subsection, we calculate the elastic displacements of an edge dislocation 
and a screw dislocation in iron. For the edge dislocation we select:
\begin{eqnarray}
 e_1''=(1,1,1)/\sqrt{3},\;e_2''=(-1,0,1)/\sqrt{2}, \; e_3''= (1,-2,1)/\sqrt{6},
\label{bcc2}
\end{eqnarray}
which are unit vectors in the directions of the Burgers vector ${\bf b}$, the normal to the 
glide plane ${\bf n}$ and the dislocation line vector, respectively. For the pure screw 
dislocation:
\begin{eqnarray}
e_1''=(-1,0,1)/\sqrt{2}, \; e_2''=(-1,2,-1)/\sqrt{6},\; e_3''= -(1,1,1)/\sqrt{3}
\label{bcc3}
\end{eqnarray}
where $e_2''$ is the normal to the glide plane and $e_3''$ a unit
vector parallel to the dislocation line and to the Burgers vector.

\begin{figure}\begin{center}
\includegraphics[width=8cm]{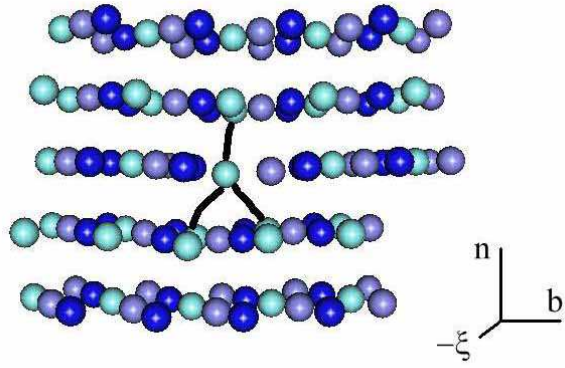}
\end{center}
\caption{(Color online) Edge dislocation in an iron lattice. Lines locating the dislocation 
core are a guide for the eye.}
\label{iron1}
\end{figure}

\begin{figure}\begin{center}
\includegraphics[width=8cm]{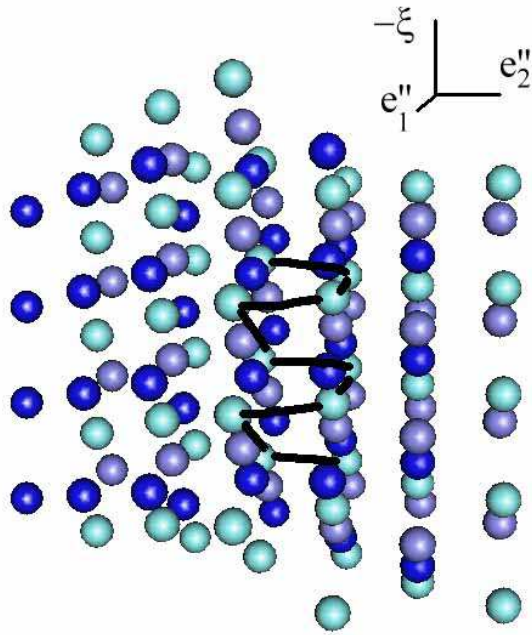}
\end{center}
\caption{(Color online) Screw dislocation in an iron lattice. Lines locating the dislocation 
core are a guide for the eye.}
\label{iron2}
\end{figure}

For iron, $C_{11}=242$ GPa, $C_{44}=112$ GPa, $C_{12}=146.5$ GPa and $H=129$ 
GPa. The lattice constant is $a= 2.87$ \AA\, and the density $\rho=7.86 $ g/cm$^3$.
Figures \ref{iron1} and \ref{iron2} show the edge and the screw dislocations obtained 
as stationary solutions of model (\ref{fcc8}). Their far fields match the corresponding 
elastic far fields of the dislocations (written in the non-orthogonal coordinates corresponding 
to the primitive cell vectors $a_1$, $a_2$, $a_3$). Dark and ligh colors are used to trace
points placed initially in different planes perpendicular to the Burgers vector.


\section{Conclusions} 
\label{sec:conclusions}
We have proposed discrete models describing defects in crystal structures whose
continuum limit is the standard linear anisotropic elasticity. The main
ingredients entering the models are the elastic stiffness constants of the
material and a dimensionless periodic function that restores the translation invariance of 
the crystal and, together with the elastic constants, determines the Peierls stress. The 
parameter value of a specific one-parameter family of periodic functions can be selected so
as to fit the observed or calculated value of the Peierls stress for the material under study. For 
simple cubic crystals, their equations of motion are derived and solved numerically
to describe simple screw and edge dislocations. Moreover, we have obtained numerically
edge dislocation loops and dipoles, and observed crack generation and growth by
applying a tension in the vertical direction to a dislocation dipole. For fcc and bcc metals, 
the primitive vectors along which the crystal is translationally invariant are not orthogonal. 
Similar discrete models and equations of motion are found by writing the strain energy density 
and the equations of motion in non-orthogonal coordinates. In these later cases, we have
determined numerically stationary edge and screw dislocations.

\section*{Acknowledegments}
We thank Ignacio Plans for a careful reading of this paper and many helpful comments. 
This work has been supported by the MCyT grant BFM2002-04127-C02, and by the 
European Union under grant HPRN-CT-2002-00282. 

\appendix
\section{Derivation of the equations of motion}
\label{appEqs}
Firstly, let us notice that 
\begin{eqnarray} 
{\partial W\over \partial u_i(l,m,n;t)} = {\partial W\over\partial e_{jk}}
{\partial e_{jk}\over \partial u_i(l,m,n;t)} = {1\over 2}\, \sigma_{jk}\,
{\partial\over \partial u_i(l,m,n;t)}\left[g(D^+_j u_k) + g(D^+_k u_j)
\right]\nonumber\\
 = {1\over 2}\, \sigma_{jk}\,\left[ g'(D^+_j u_k) {\partial (D^+_j u_k)
\over \partial u_i(l,m,n;t)}+ g'(D^+_k u_j) {\partial (D^+_k u_j)\over \partial
u_i(l,m,n;t)}\right] , \label{B1}
\end{eqnarray} 
where $W$ is a function of the point $(l',m',n')$, and we have used the
definition of stress tensor:
\begin{eqnarray}
\sigma_{ij} = {\partial W\over\partial e_{ij}} , \label{B2}
\end{eqnarray} 
and its symmetry, $\sigma_{ij} = \sigma_{ji}$. Now, we have
\begin{eqnarray} 
{\partial \over \partial u_i(l,m,n;t) }[D^+_1 u_k(l',m',n';t)] =\delta_{ik}
\, (\delta_{l\, l'+1} - \delta_{ll'})\, \delta_{m m'} \delta_{n n'} ,
\label{B3}
\end{eqnarray} 
and similar expressions for $j= 2,3$. By using (\ref{B1}) - (\ref{B3}), we
obtain
\begin{eqnarray} {\partial\over \partial u_i(l,m,n;t)} \sum_{l',m',n'}
W(l',m',n';t) = - \sum_j D^-_j  [\sigma_{ij}\,g'(D^+_j u_i)]
\,. \label{B4}
\end{eqnarray} 
In this expression, no sum is intended over the subscript $i$, so that we have abandoned 
the Einstein convention and explicitly included a sum over $j$. Therefore Eq.\ (\ref{e11}) 
for conservative dynamics becomes
\begin{eqnarray} 
M\, \ddot{u}_{i} &=& \sum_j D^-_j [\sigma_{ij}\,g'(D^+_j u_i)],
\label{B5}
\end{eqnarray}
which yields Eq.\ (\ref{e13}). Except for the factor $g'(D^+_j u_i)$, these equations 
are discretized versions of the usual ones in elasticity \cite{ll7}.

\end{document}